\documentclass[11pt]{article}
\usepackage{amssymb}
\usepackage{amsmath}
\usepackage{graphicx}
\renewcommand{\theequation}{\arabic{section}.\arabic{equation}}
\def\lb{\label}
\def\bb{\bibitem}
\def\be{\begin{equation}}
\def\ee{\end{equation}}
\def\ba{\begin{eqnarray}}
\def\ea{\end{eqnarray}}
\def\dfrac{\displaystyle\frac}
\def\nn{\nonumber}
\def\ol{\overline}

\def\S{\mathbf{S}}
\def\X{\mathbf{X}}
\def\V{\mathbf{V}}
\def\a{\boldsymbol{\alpha}}
\def\b{\boldsymbol{\beta}}
\def\c{\boldsymbol{\gamma}}
\def\x{\boldsymbol{\xi}}
\def\R{{\cal R}}
\def\k{\kappa}
\def\J{\mathbf{J}}

\def\rq{\rho_0^{2q-1}}
\def\5{_{(5)}}

\begin{document}
\begin{titlepage}
\date{24 October 2015}

\title{\begin{flushright}\begin{small}    LAPTH-058/15
\end{small} \end{flushright} \vspace{1.5cm}
Chern-Simons dilaton black holes in 2+1 dimensions}

\author{Karim Ait Moussa$^a$ \thanks{Email: k.aitmoussa@yahoo.fr},
G\'erard Cl\'ement$^b$ \thanks{Email: gclement@lapth.cnrs.fr}, Hakim
Guennoune$^{b,c}$ \thanks{Email: guenounnehakim@yahoo.fr} \\ \\
{$^a$ \small Laboratoire de Physique Math\'ematique et Physique
Subatomique, D\'epartement de Physique,} \\{\small Facult\'e des
Sciences, Universit\'e Mentouri, Constantine 25000, Algeria} \\
{$^b$ \small LAPTh, Universit\'e Savoie Mont Blanc, CNRS, 9 chemin
de Bellevue,} \\{\small B.P.110, F-74941 Annecy-le-Vieux cedex, France}\\
{$^c$ \small D\'epartement de Physique, Facult\'e des Sciences,
Universit\'e Ferhat Abbas, S\'etif 19000, Algeria}}
\maketitle

\begin{abstract}
We construct rotating magnetic solutions to the three-dimensional
Einstein-Maxwell-Chern-Simons-dilaton theory with a Liouville
potential. These include a class of black hole solutions which
generalize the warped AdS black holes. The regular black holes
belong to two disjoint sectors. The first sector includes black
holes which have a positive mass and are co-rotating, while the
black holes of the second sector have a negative mass and are
counter-rotating. We also show that a particular, non-black hole,
subfamily of our three-dimensional solutions may be uplifted to new
regular non-asymptotically flat solutions of five-dimensional
Einstein-Maxwell-Chern-Simons theory.
\end{abstract}

\end{titlepage}

\setcounter{page}{2}

\section{Introduction}
Lower dimensional gravity provides an arena for constructing exact
analytical black hole solutions in a wide variety of gravitating
field theories. Most of the known black hole solutions are static.
The celebrated BTZ black hole in $2+1$ dimensions \cite{btz} can be
either static or rotating, but the rotating solutions can be
transformed to the static one by a local coordinate transformation.
The first intrisincally rotating black holes, now known as warped
AdS$_3$ black holes \cite{ALPSS}, were constructed as solutions to
topologically massive gravity \cite{djt} in \cite{tmgbh}, and then
generalized to solutions of cosmological topologically massive
gravity in \cite{BC}, of topologically massive
gravitoelectrodynamics (TMGE) \cite{tmge}  in \cite{ACGL}, of new
massive gravity \cite{bht} in \cite{newmass}, of $R^3$ extended new
massive gravity \cite{sinha} in \cite{emg}, and of generalized
massive gravity \cite{bht} in \cite{gmg}.

In this paper, we shall extend the warped AdS black hole class to a
more general class of intrinsincally rotating black holes. The
warped AdS black holes of \cite{ACGL} were solutions of
three-dimensional Einstein-Maxwell theory augmented by both
gravitational and electromagnetic Chern-Simons terms \cite{tmge},
which reduces to Einstein-Maxwell-Chern-Simons theory when the
gravitational Chern-Simons term is absent. In \cite{CF1}, a class of
rotating electric solutions to the cosmological
Einstein-Maxwell-Chern-Simons-dilaton theory were obtained for a
special relation between the model parameters. From these solutions,
a class of rotating magnetic solutions to the same theory were
generated in \cite{CF2}. Considering this theory, we discuss in the
next section the reduction of the field equations along the lines
followed in \cite{ACGL}. We then show in Sect. 3 that a simple
ansatz generalizing that of \cite{ACGL} leads to rotating magnetic
solutions more general than those of \cite{CF2}. The global
structure of these solutions is analyzed in Sect. 4, where a
subclass of regular black hole solutions is found in a certain
parameter range. The mass, angular momentum and other observables of
these black holes are computed in Sect. 5. In Sect. 6 we show that
another subclass of our solutions may be uplifted to new regular
non-asymptotically flat solutions of five-dimensional
Einstein-Maxwell-Chern-Simons theory. Our results are summarized in
the final section.

\section{Reduction of the field equations}
\setcounter{equation}{0}

Self-gravitating dilatonic topologically massive electrodynamics is
defined by \cite{CF1}
\begin{eqnarray}\lb{act}
I_{E} &=&\int
d^{3}x\sqrt{|g|}\left[\frac1{2\kappa}R-2g^{\mu\nu}\partial_{\mu}\phi
\partial_{\nu}\phi - \frac{\Lambda}{\kappa}e^{b\phi} -
\frac{1}{4}e^{c\phi}g^{\mu\nu}g^{\rho\sigma}F_{\mu\rho}F_{\nu\sigma}
\right] \nonumber\\ && + \frac{\mu}{2}\int
d^{3}x\epsilon^{\mu\nu\rho}A_{\mu}\partial_{\nu}A_{\rho}\,,
\end{eqnarray}
with $\kappa =8\pi G$ the Einstein gravitational constant (which in
$2+1$ dimensions can be either positive or negative), $b$ and $c$
the coupling constants of the dilatonic field $\phi$ to the
cosmological constant $\Lambda $ and to the Maxwell field $F=dA$,
and $\mu$ the Chern-Simons coupling constant
($\epsilon^{\mu\nu\rho}$ is the antisymmetric symbol).

Assuming the existence of two commuting Killing vectors, we choose
the parametrisation \cite{C93,tmge}
\begin{eqnarray}
ds^{2} &=&\lambda _{ab}\left( \rho \right) dx^{a}dx^{b}+\zeta
^{-2}\left( \rho \right) R^{-2}\left( \rho \right) d\rho ^{2} \quad
(a,b=0,1), \nonumber \\
A &=&\psi _{a}\left( \rho \right) dx^{a}    \label{para}
\end{eqnarray}%
($x^{0}=t,$ $x^{1}=\varphi $), where $\lambda $ is the $2\times 2$
matrix
\begin{equation}
\lambda =\left(
\begin{array}{cc} T+X & Y \\
Y & T-X
\end{array}
\right)  \label{lambda}\,,
\end{equation}
$R^{2}\equiv \mathbf{X}^{2}$ is the Minkowski pseudo-norm of the
vector $\mathbf{X}\left( \rho \right) =\left( T,X,Y\right) $
\begin{equation}\label{mink}
\mathbf{X}^{2}=\eta _{ij}X^{i}X^{j}=-T^{2}+X^{2}+Y^{2}\,,
\end{equation}
and $\zeta$ a scale factor. The stationary sector of the spacetime
of metric (\ref{para}) corresponds to the spacelike sector $R^2>0$
of (\ref{mink}), with the same signature $(-++)$.

This parametrisation reduces the action (\ref{act}) to the form
\begin{equation}
I=\int d^{2}x\int d\rho\, L
\end{equation}
where $L$ is the effective Lagrangian
\begin{equation}
L=\frac{1}{2}\left[ \frac{\zeta}{2\kappa}\mathbf{X}^{\prime 2} -
4\zeta R^{2}\phi^{\prime 2} - \frac{2\Lambda}{\kappa\zeta} e^{b\phi
}  + \zeta e^{c\phi}\overline{ \psi}^{\prime}\left(\mathbf{\Sigma} .
\mathbf{X}\right)\psi^{\prime} + \mu\overline{\psi}\psi^{\prime}
\right]  \label{lag}\,,
\end{equation}
where $^{\prime}= \partial/\partial\rho$, the "Dirac" matrices
$\Sigma
 ^{i}$ are defined by
\begin{equation}
\Sigma ^{0}=\left(
\begin{array}{cc}
0 & 1 \\
-1 & 0%
\end{array}
\right) \ ,\ \Sigma ^{1}=\left(
\begin{array}{cc}
0 & -1 \\
-1 & 0
\end{array}
\right) \ ,\ \Sigma ^{2}=\left(
\begin{array}{cc}
1 & 0 \\
0 & -1
\end{array}
\right)\,,  \label{sig}
\end{equation}
and $\overline{\psi }\equiv \psi ^{T}\Sigma ^{0}$ is the Dirac
adjoint of the "spinor" $\psi$. In passing from (\ref{act}) to
(\ref{lag}), we have set the sign convention for the antisymmetric
symbol to $\epsilon^{012}=-1$ (corresponding to $\epsilon
_{012}=+1$).

Variation of the Lagrangian $L$ with respect to $\psi$ gives the
equation
\begin{equation}
\partial_{\rho }\left[ \mu\psi - \zeta e^{c\phi }\left( \mathbf{\Sigma
.X}\right) \psi ^{\prime }\right] =0\,, \label{eqpsi}
\end{equation}
which is integrated, up to a gauge transformation, by
\begin{equation}\lb{firstpsi}
\zeta \psi ^{\prime }=\frac{\mu}{R^{2}}e^{-c\phi}( \mathbf{\Sigma
.X})\psi \,.
\end{equation}
Defining the wedge product of two vectors $\X$ and $\mathbf Y$ by
 \be
\left( \mathbf{X\wedge Y}\right)^{i}=\eta^{ij}\epsilon
_{jkl}X^{k}Y^{l}
 \ee
with $\epsilon _{012}=+1$, the spinor equation (\ref{firstpsi}) can
be shown \cite{tmge} to be equivalent to the vector dynamical
equation
\begin{equation}
\zeta \mathbf{S}^{\prime }=\frac{2\mu}{R^{2}}e^{-c\phi }\mathbf{X}
\wedge \S \label{eqS}
\end{equation}
for  the ``spin'' vector
\begin{equation}
\S=-\frac{\kappa }{2}\overline{\psi }\mathbf{\Sigma }\psi
\label{defS}
\end{equation}
of components
 \be\lb{compS}
S^0 = \frac{\kappa}2\left[\psi_0^2 + \psi_1^2\right]\,, \quad S^1 =
\frac{\kappa}2\left[\psi_0^2 - \psi_1^2\right]\,, \quad S^2 =
\kappa\psi_0\psi_1\,.
 \ee
Note that our trading of a spinor equation (\ref{firstpsi}) for a
vector equation (\ref{eqS}) is possible only because the spin vector
is null,
 \be\lb{Snull}
\S^2 = 0.
 \ee
Next, vary $L$ with respect to $\mathbf{X}$, leading to the equation
\begin{equation}
\V \equiv \mathbf{X}^{\prime\prime} + 8\kappa\mathbf{X}\phi^{\prime
2} = \kappa
e^{c\phi}\overline{\psi}^{\prime}\mathbf{\Sigma}\psi^{\prime}
 = \frac{2\mu^{2}}{\zeta^2R^{2}}e^{-c\phi }\left[
\frac{2}{R^{2}}\left(\S.\mathbf{X}\right) \mathbf{X}-\S\right]
\label{eqX}\,,
\end{equation}
where we have used Eq. (\ref{firstpsi}) and the definition
(\ref{defS}). Similarly, variation of $L$ with respect to $\phi$
leads to
\begin{equation}
\left(R^{2}\phi^{\prime}\right)^{\prime} =
\frac{b\Lambda}{4\kappa\zeta^2}e^{b\phi }
-\frac{c\mu^{2}}{4\kappa\zeta^{2}R^{2}}e^{-c\phi}\left(\S.\mathbf{X}\right)\,.
\label{eqphi}
\end{equation}
Finally, variation of the Lagrangian with respect to the Lagrange
multiplier $\zeta$ leads to the Hamiltonian constraint
\begin{equation}
H\equiv \frac{1}{4\kappa }\left[\mathbf{X}^{\prime 2}-8\kappa
R^{2}\phi^{\prime 2} + \frac{4\Lambda}{\zeta^{2}} e^{b\phi} +
\frac{4\mu^{2}}{\zeta^{2}R^{2}}e^{-c\phi}\left(\S.\mathbf{X}\right)
\right] = 0 \label{ham}\,.
\end{equation}

Equations (\ref{eqphi}) and (\ref{ham}) involve the scalar $\S.\X$
which can be evaluated by taking the scalar product of Eq.
(\ref{eqX}) with $\mathbf{X}$ as
\begin{equation}
\left( \S.\mathbf{X}\right) =\frac{R^{2}\zeta ^{2}}{2\mu^{2}%
}e^{c\phi }\left( \mathbf{X.X}^{^{\prime \prime }}+8\kappa R^{2}\phi
^{\prime 2}\right) \,. \label{SX}
\end{equation}
Inserting this in Eqs. (\ref{eqphi}) and (\ref{ham}) we obtain the
simplified equations
\begin{equation}\lb{eqphi1}
\left( R^{2}\phi ^{\prime }\right) ^{\prime }+\frac{c}{8\kappa
}\left(
\mathbf{X.X}^{^{\prime \prime }}+8\kappa R^{2}\phi ^{\prime 2}\right) -\frac{%
b\Lambda }{4\kappa \zeta ^{2}}e^{b\phi }=0
\end{equation}
and
\begin{equation}
\mathbf{X}^{\prime 2}+2\mathbf{X.X}^{^{\prime \prime }}+8\kappa
R^{2}\phi ^{\prime 2}+\frac{4\Lambda }{\zeta ^{2}}e^{b\phi }=0\,.
\label{ham1}
\end{equation}
These last two equations may be combined to yield
\begin{equation}
\left( R^{2}\phi^{\prime}-\frac{c}{8\kappa}\left(
\mathbf{X.X}^{\prime}\right)\right)^{\prime} =
\frac{\Lambda}{2\kappa\zeta^{2}}\left(c+\frac{b}{2}\right)e^{b\phi}
\label{mast}\,.
\end{equation}
This last equation can be easily integrated only if the dilaton
coupling constants are related by
$$b+2c=0\,,$$
which we assume henceforth. This integration leads to
\begin{equation}\lb{mast1}
R^{2}\phi^{\prime} = \frac{c}{8\kappa }\left(
RR^{\prime }+d\right)\,,
\end{equation}
where $d$ is an integration constant.

Finally, the independent reduced field equations can be taken to be
(\ref{mast1}), (\ref{ham1}) and (\ref{eqS}) (where the spin vector
$\S$ is given in terms of $\X$ and $\phi$ by (\ref{eqX})),
supplemented by the conditions, following from (\ref{Snull}) and
(\ref{compS}), that the vector $\kappa^{-1}\S$ is future null,
 \be
\S^2 = 0\,, \quad \kappa^{-1}S^0 > 0\,.
 \ee

\setcounter{equation}{0}
\section{Black Hole Solutions}

\subsection{Power-law ansatz}
Black hole solutions were found in TMGE by making the ansatz \cite{ACGL}
 \be
\X = \a\rho^2 + \b\rho + \c\,,
 \ee
where $\a$, $\b$ and $\c$ are three linearly independent constant
vectors, and the vector $\a$ is null and orthogonal to the vector
$\b$,
\begin{equation}\lb{anull}
\a^{2}=0\,, \quad \a.\b = 0\,.
\end{equation}

In the present dilatonic context, we shall generalize this ansatz by assuming
\begin{equation}
\mathbf{X}= \a\rho^{q+p} + \b\rho^{q}+ \c\rho^{q-p}\,, \label{an}
\end{equation}
where $q$ and $p\neq0$ are real numbers (the ansatz (\ref{an})
degenerates for $p=0$) which should reduce to $q=p=1$ in the limit
$c=0$ of vanishing dilaton coupling, and the constant vectors $\a$,
$\b$ and $\c$ are again constrained by (\ref{anull}). It follows from these assumptions that
 \be\lb{R2}
R^{2} = m\rho ^{2q} + n\rho^{2q-p} + l\rho ^{2q-2p}\,,
 \ee
with
 \be
m = 2\a.\c + \b^2\,, \quad n = 2\b.\c\,, \quad l = \c^2\,.
 \ee
Inserting the functional form (\ref{R2}) of $R^2(\rho)$ into
(\ref{mast1}), we obtain
\begin{equation}\lb{phiprime1}
\phi ^{\prime }=\frac{c}{16\kappa }\frac{2qm\rho ^{2q-1}+ (2q-p)n\rho ^{2q-p-1}
+ (2q-2p)l\rho ^{2q-2p-1} + 2d}{m\rho ^{2q}+n\rho^{2q-p} +l\rho ^{2q-2p}}\,.
\end{equation}
This can be easily integrated to $\phi \propto \ln\rho$ if 1) the constant $2d$
can be grouped with one of the three monomials in the numerator, implying either
$2q-1=0$, $2q-p-1=0$, or $2q-2p-1=0$, only the second possibility
 \be\lb{pq}
p = 2q-1
 \ee
being consistent with the constraint that $p$ and $q$ reduce to $1$ in the limit
of vanishing dilaton coupling; 2) the coefficients in the numerator are
matched to those in the denominator so that $\phi' \propto \rho^{-1}$:
 \be
n + 2d = 2qn\,, \quad 2(1-q)l = 2ql\,.
 \ee
(where we have replaced $p$ in terms of $q$ according to (\ref{pq})). The first
relation fixes the integration constant $d$ for the dilaton field
equation (\ref{eqphi1}) in terms of the metric parameters. The second relation
is solved either by $q=1/2$, leading to $p=0$, which we have excluded, or by $l=0$. We are thus
led to add to the ansatz (\ref{an}), (\ref{anull}) the complementary assumption
 \be\lb{cnull}
\c^2 = 0\,,
 \ee
leading to the solution of (\ref{phiprime1})
 \be
\phi = \frac{cq}{8\kappa}\,\ln\left(\frac{\rho}{\rho_1}\right)\,,
 \ee
with $\rho_1>0$ a new integration constant.

Next, fixing without loss of generality the scale parameter $\zeta$
to
 \be
\zeta=\mu\,,
 \ee
we compute the left-hand side of Eq. (\ref{eqX}):
 \be
\V = \left[(3q-1)(3q-2)+\frac{c^2q^2}{8\kappa}\right]\a\rho^{3q-3} +
\left[q(q-1)+\frac{c^2q^2}{8\kappa}\right]\left(\b\rho^{q-2} +
\c\rho^{-q-1}\right)\,.
 \ee
Squaring the right-hand side of Eq. (\ref{eqX}), we find that this
vector must be null, as the spin vector $\S$ (Eq. (\ref{Snull})).
This is ensured if $\V$ is collinear with $\a$,
which is possible only if
\begin{equation}\label{valq}
q=\frac{1}{1+c^2/8\kappa}
\end{equation}
(the other possibility $q=0$ does not lead to a consistent
solution). It follows that $c^2q/8\kappa=1-q$, so that
 \be\lb{ecphi}
e^{c\phi} = \left(\frac{\rho}{\rho_1}\right)^{1-q}\,, \quad \V = 2(1-2q)^2\a\,
\rho^{3q-3}\,.
 \ee
Computing the spin vector from the inverse of (\ref{eqX}),
 \be
\S = \frac12\,e^{c\phi}\left[2(\V.\X)\X - R^2\V\right]\,,
 \ee
we obtain
 \ba\lb{Sabc}
\S &=& (1-2q)^2\rho_1^{q-1}\left[2(\a.\c)\X\rho^{q-1} -
R^2\a\rho^{2q-2}\right] \nn \\
&=& (1-2q)^2\rho_1^{q-1}\left[k\a\wedge\b\,\rho^{4q-2}
+2k\a\wedge\c\,\rho^{2q-1} + 2(\a.\c)\c\right]\,.
 \ea
To obtain the second form of $\S$, we have used (\ref{R2}), and
the vector relations, which follow from (\ref{anull})
 \be\lb{k}
\a \wedge \b = -k\a\,, \quad \b^2 = k^2\,,
 \ee
where $k$ is some constant. We then see that the dynamical equation (\ref{eqS})
is satisfied provided
 \be\lb{kq}
k = \frac{\rho_1^{1-q}}{2q-1}\,.
 \ee
Finally, the Hamiltonian constraint in the form (\ref{ham1}) is also
satisfied provided
 \be\lb{valz}
\a.\c = \frac{q + 4(2q-1)\Lambda/\mu^2}{2(1-3q)}\,k^2\,,
 \ee
for $q \neq 1/3$. For $q=1/3$ ($c^2 = 16\kappa$) and $\Lambda=\mu^2/4$,
the value of the scalar product $\a.\c$ remains arbitrary.

We choose for the basis vectors $\a$, $\b$ and $\c$ the parametrisation, which
generalizes that made in \cite{ACGL},
\begin{equation}\lb{abc}
\a = k^3\left(\epsilon/2,-\epsilon/2,0\right)\,, \quad \b =
k\left(\omega ,-\omega ,-1\right)\,, \quad \c =
k^{-1}\left(z+u,z-u,v\right)
\end{equation}
($\epsilon^2 = 1$). These automatically satisfy (\ref{anull}) and (\ref{k}),
while the constraint (\ref{cnull}) implies the relation
 \be\lb{uvz}
u = v^2/4z\,.
 \ee
The value of the parameter $z$
may be computed from $\a.\c = -\epsilon k^2z$, where the scalar product $\a.\c$
is given by (\ref{valz}). The scalar function $R^2$ is, from (\ref{R2}),
 \be
R^{2} = k^2\beta^{2}\rho \left(\rho^{2q-1}-\rho_0^{2q-1}\right)\,,
 \ee
where the parameters $\beta^2$ (real) and $\rq$ ($\rho_0>0$) are
defined by
 \ba\lb{paras}
k^2\beta^{2} &=& k^2(1-2\epsilon z) = \b^2 + 2\a.\c\,, \nn\\
k^2\beta^{2}\rho _{0}^{2q-1} &=& 2(v+2\omega z) = -2\b.\c\,,
 \ea
leading to
 \be
\omega = \frac{\epsilon}{2(1-\beta^2)}\left(k^2\beta^2\rq
-2v\right)\,.
 \ee
 The resulting metric may be written in terms of the three
parameters $k$, $\rho_0$ and $v$ as
\begin{eqnarray}
ds^{2} &=& \frac{\epsilon(1-\beta^{2})}k\,\rho ^{1-q}\left[dt-
\frac{\epsilon(k^2\rho ^{2q-1}-v)}{1-\beta^{2}}d\varphi\right]^{2} \nonumber \\
&& -\frac{\epsilon k^3\beta ^{2}\rho^q(\rho ^{2q-1}-\rho_0
^{2q-1})}{1-\beta^{2}} d\varphi^{2}
+\frac{d\rho^{2}}{k^2\mu^{2}\beta
^{2}\rho\left[\rho^{2q-1}-\rho_0^{2q-1} \right]}\,,  \label{genmet}
\end{eqnarray}
with $\beta^2$ a real parameter given in terms of the model
parameters by
\begin{equation}\lb{valb2}
\beta^2 =\frac{1-2q}{1-3q}\left(1-\frac{4\Lambda}{\mu^{2}}\right)\,,
\end{equation}
when $q\neq1/3$, $\beta^2$ remaining arbitrary when $q=1/3$ with
$\Lambda=\mu^2/4$. Solving Eq. (\ref{compS}) for $\psi$, we obtain
the electromagnetic field generating this gravitational field:
\begin{equation}\lb{genA}
A=\pm\sqrt{\frac{\epsilon(1-2q)}{\kappa}}\left[\epsilon(1-\beta^{2})dt
- (k^2\rho ^{2q-1}-v)d\varphi\right]\,.
\end{equation}
The constant electric potential is irrelevant and may be gauged away, the
magnetic potential $A_{\varphi}$ leading both to a magnetic field and to an
electric field $F^{t\rho}$ because the metric (\ref{genmet}) is non-diagonal.
This electromagnetic field is real provided
 \be\lb{Areal}
\epsilon = {\rm sign}[\kappa(1-2q)]\,.
 \ee
For $q=1$ ($c=0$), the solution (\ref{genmet}), (\ref{genA}) reduces
to the non-dilatonic solution obtained in \cite{ACGL} for
$\lambda=0$ (no gravitational Chern-Simons term). Note however that
the black holes of \cite{ACGL} are regular for $0<\beta^2<1$, while
as we shall see in the next section the present dilatonic solution
can correspond to regular black holes only if $\beta^2<0$.

The form of the metric (\ref{genmet}) breaks down for $\beta^2=1$
and $\beta^2=0$. For $\beta^2=1$ ($\Lambda=(q/(1-2q))\mu^2/4$),
$z=0$ implying $v = 0$ from (\ref{uvz}), so also $\rho_0=0$. The
metric, depending on the two independent parameters $\omega$ and
$u$, may be written in the ADM form (\ref{adm}) with
$v^2/(1-\beta^2)$ replaced with $2\epsilon u$. For $\beta^2=0$
($\Lambda=\mu^2/4$ with $q\neq1/3$), the metric is obtained from
(\ref{genmet}) by replacing $-k^2\beta^{2}\rho _{0}^{2q-1}$ with
$2\nu\equiv-2(v+\epsilon\omega)\neq0$.

To conclude this part, we comment on the relation of our magnetic power-law
solution with that of Castelo Ferreira \cite{CF2} (similar comments can be
made concerning the electric solution of \cite{CF1}). The space-metric
metric of \cite{CF2} is parametrized by
 \be
ds^2 = -f^2(dt+A\,d\varphi)^2 + dr^2 + h^2\,d\varphi^2\,,
 \ee
with $f^2 \propto r$, $h^2 \propto  r^{2p_{CF}-1}$, $p_{CF} = 4\kappa/c^2$. This
metric can be transformed to the form (\ref{para}) by the radial coordinate
transformation $d\rho = \mu R dr$, with $R^2 = f^2h^2 \propto  r^{2p_{CF}}$,
leading to $\rho \propto r^{p_{CF}+1}$. The resulting vector $\X(\rho)$ is
of the form (\ref{an}) with $p = (p_{CF}-1)/(p_{CF}+1)$ and
$q = p_{CF}/(p_{CF}+1)$, different from our exponents (\ref{pq}) and (\ref{valq}),
the basis vectors $\a$, $\b$ and $\c$ obeying the constraints (\ref{anull}) and
(\ref{cnull}), together with the additional constraint
 \be\lb{n0}
\b.\c = 0\,,
 \ee
accounting for the fact that $R^2 \propto \rho^{2q}$ is a monomial.
The price to pay for this additional constraint is that the
Hamiltonian constraint (\ref{ham}) can only be satisfied if there is
a specific relation between the dimensionless coupling constants
$c^2/\kappa$ and $\Lambda/\mu^2$, Eq. (3.6) of \cite{CF2}.

\subsection{Case $c^{2}=8\kappa$}

For $c^{2}=8\kappa$, (\ref{valq}) leads to $q=1/2$, corresponding to
$p=0$ from (\ref{pq}), so that the power-law ansatz (\ref{an}) is degenerate.
In this case, we replace this ansatz by the logarithmic ansatz
\begin{equation}
\mathbf{X}= \a\rho^{1/2}\ln(\rho/\rho_0) + \b\rho^{1/2}\,,
\end{equation}
with only two linearly independant constant vectors $\a$ and $\b$ obeying
the constraints (\ref{anull}). This leads to $R^2 = m\rho$, so that from
(\ref{mast1}),
 \be
\phi' = \frac{m+2d}{2cm\rho}\,.
 \ee
The computation of $\V$ then leads to
 \be
\V = \frac{d(m+d)}{m^2}\,\rho^{-2}\X\,,
 \ee
which is null, $\V^2=0$, only if $d(m+d)=0$, leading to
 \be
\phi' = \pm \frac1{2c\rho}\,, \quad \V = 0\,,
 \ee
implying also $\S = 0$, which in turn means $\psi=0$, and
 \be
\phi = \pm\frac{1}{2c}\ln(4\rho/a)\,,
 \ee
with $a>0$ an integration constant. Finally, insertion into the
Hamiltonian constraint (\ref{ham}) shows that it can be satisfied
only if the cosmological constant vanishes, $\Lambda=0$.

Choosing the parametrisation
 \be
\a = a^{-1/2}(1/2,\; 1/2,\; 0)\,, \quad \b =
a^{-1/2}(0,\; 0,\; 1)\,,
 \ee
and taking in (\ref{para}) $x^0=v$, $x^1=u$, $\zeta=1$, we obtain the
metric
\begin{equation}
ds^{2}= -2\left(\frac{\rho}{a}\right)^{1/2}dudv +
\left(\frac{\rho}{a}\right)^{1/2}\ln(\rho/\rho_0)\,du^2
+\frac{ad\rho ^{2}}{\rho }\,.
\end{equation}
This may be simplified by the radial coordinate transformation $\rho
=ax^{2}/4$ ($x>0$), leading to the solution
\begin{eqnarray}\lb{sollog}
ds^{2} &=&  - x\,dudv + x\ln(x/x_0)\,du^2 + a^2\,dx^{2}\,, \nn  \\
e^{c\phi} &=& x^{\pm1}\,, \qquad A = 0\,.
\end{eqnarray}

This is a solution of three-dimensional gravity without
cosmological constant minimally coupled to a scalar field. It can be
related to known solutions of four-dimensional vacuum gravity in the
following manner. Assuming the existence of a spacelike Killing
vector $\partial_y$, the Kaluza-Klein reduction ansatz
\begin{equation}
ds_{(4)}^{2}=e^{c\phi }dy^{2}+e^{-c\phi }ds_{(3)}^{2}  \label{(3.36)}
\end{equation}
reduces the four-dimensional Einstein-Hilbert action to the
three-dimensional action (\ref{act}) with $A = 0$, $\Lambda=0$ and
$2\kappa=c^2/4$. Conversely, the lift of the three-dimensional
solution (\ref{sollog}) to four dimensions yields the two solutions
\begin{equation}\lb{4+}
ds_{(4)+}^{2}=xdy^{2} - dudv + \ln(x/x_0)\,du^{2} + a^2x^{-1}\,dx^{2}\,,
\end{equation}
\begin{equation}\lb{4-}
ds_{(4)-}^{2} = x^{-1}dy^{2} -x^{2}\,dudv + x^{2}\ln(x/x_0)\,du^{2} +
a^2x\,dx^{2}\,.
\end{equation}
The metric (\ref{4+}) can be transformed into the special $pp$-wave solution
\cite{EK62}
\begin{equation}
ds_{(4)+}^{2}=d\zeta d\overline{\zeta }-dudv-2{\rm Re}\,f(\zeta)\,du^{2}\,
\end{equation}
with
\begin{equation}
\zeta = 2ax^{1/2}e^{iy/2a}\,, \quad f(\zeta) = -\ln(\zeta/2ax_0^{1/2})\,,
\end{equation}
while the transformation $x = (\xi/a)^{1/2}$ puts the metric (\ref{4-}) into
the form
\begin{equation}
ds_{(4)-}^{2}= \left(\frac{\xi}a\right)^{-1/2}(d\xi^{2}+dy^{2})
-\frac{\xi}a\,dudv + \frac{\xi}{2a}\ln(\xi/ax_0^2)\,du^2
\end{equation}
of a special van Stockum solution \cite{vanS}.

\setcounter{equation}{0}
\section{Regularity and global structure}
In this section, we investigate under which conditions the metric
(\ref{genmet}) is regular. This metric, written in
Arnowitt-Deser-Misner (ADM) form as:
 \ba\lb{adm}
ds^2 &=& -\frac{\beta^2\rho(\rho^{2q-1}-\rho_0^{2q-1})}{r^2}\,dt^2 +
r^2\bigg[d\varphi
  - \epsilon\epsilon'\,\frac{\rho^q-v\rho^{1-q}}{r^2}\,dt\bigg]^2 \nn \\ &&\qquad
+ \frac{d\rho^2}{\mu^2\beta^2\rho(\rho^{2q-1}-\rho_0^{2q-1})}\,,
 \ea
with
 \be\lb{r2}
r^2 = \epsilon'\left[ \rho^{3q-1} +2\epsilon\omega\rho^q +
\frac{v^2}{1-\beta^2}\,\rho^{1-q} \right]\,,
 \ee
depends on two arbitrary parameters $\rho_0$ and $v$, to which
$\omega$ is related by (\ref{paras}). In (\ref{adm}), the real
parameters $\beta^2$ and $q$ are related to the model parameters by
(\ref{valb2}) and (\ref{valq}), and we have normalized the arbitrary
scale $k$ defined in (\ref{kq}) so that $k^2=1$. From Eq.
(\ref{kq}), the sign of $k$ is that of $(2q-1)$, so that the
condition (\ref{Areal}) for the reality of the electromagnetic field
leads to $k = \epsilon\epsilon'$, with
 \be\lb{epskq}
\epsilon' = - {\rm sign}(\kappa)\,, \quad \epsilon\epsilon' = {\rm
sign}(2q-1)\,.
 \ee

The associated Ricci scalar is
 \be\lb{riscal}
\R = \frac{\mu^2}2\left\{[(2q-1)^2 -
(11q^2-8q+1)\beta^2]\,\rho^{2q-2} +
q(q-1)\beta^2\rho_0^{2q-1}\rho^{-1}\right\}\,.
 \ee
For $\beta^2\neq0$ this diverges, whatever the value of $q$, for
$\rho\to0$ if $\rho_0^{2q-1}\neq0$. The corresponding curvature
singularity is safely hidden behind the horizon if $\rho_0^{2q-1}>0$
and if $\rho\to\infty$ corresponds to spacelike infinity. For
$\beta^2=0$, the constant $\beta^{2}\rho _{0}^{2q-1}$ is replaced
with $-2\nu\neq0$, so that there is always a naked curvature
singularity at $\rho=0$. Eq. (\ref{riscal}) shows that there is also
generically a curvature singularity at $\rho\to\infty$ if $q
> 1$, so we must restrain $q$ to the range $q<1$ for regularity (we
exclude the value $q=1$, which corresponds to the case of TMGE
treated in \cite{ACGL}).

We now determine the parameter ranges for which the metric has the
Lorentzian signature $(-++)$, implying $r^2=T-X>0$ and $g_{\rho\rho}
= \mu^{-2}R^{-2} > 0$, outside the horizon $\rho=\rho_0$. This
signature depends on the signs of $\epsilon$ and $\epsilon'$, which
depend on the sign of $\kappa$ and the value of $q$ through
(\ref{epskq}). It also follows from the relation (\ref{valq})
defining the exponent $q$ that
 \ba
{\rm for} & \kappa > 0 \quad (\epsilon' = -1)\,, & \quad  0 < q < 1\,, \nn\\
{\rm for} & \kappa < 0 \quad (\epsilon' = +1)\,, & \quad  q < 0\,.
 \ea

We first consider the case $\kappa > 0$ ($\epsilon'=-1$). If $q >
1/2$ ($\epsilon=-1$), $r^2$ is dominated for large $\rho$ by the
first term of (\ref{r2}), $\epsilon'\rho^{3q-1}$, which is negative,
so that there are closed timelike curves (CTC) at infinity. If $q <
1/2$ ($\epsilon=+1$), $R^2$ is positive for large $\rho$ provided
$\beta^2<0$, while if $v\neq0$, $r^2$ is now dominated for large
$\rho$ by the last term of (\ref{r2}), $\epsilon'
v^2\rho^{1-q}/(1-\beta^2)$, which is again negative, leading again
to CTC at infinity. On the other hand, if $q < 1/2$ and $v=0$, $r^2$
is dominated for large $\rho$ by the middle term
 \be\lb{midv0}
2\epsilon\epsilon'\omega\rho^q =
\frac{-\beta^2\rho_0^{2q-1}}{1-\beta^2}\rho^q\,,
 \ee
which is positive. But the negative first term gains importance when
$\rho$ decreases, so that $r^2 = T - X$ vanishes at some value
$\rho_c$. It then follows from (\ref{mink}) that $R^2(\rho_{c}) =
Y^2(\rho_c) \ge0$, so that $\rho_{c}\ge\rho_0$, i.e. the causal
singularity $\rho=\rho_c$ is naked. The conclusion is that there are
no regular black holes for $\kappa > 0$.

Now we consider the case $\kappa < 0$ ($q<0$, $\epsilon'=1$,
$\epsilon=-1$). Again, $R^2$ is positive provided $\beta^2<0$ but,
for $v\neq0$, the last term (dominant for large $\rho$) of $r^2$ is
now positive. We show in Appendix A that $r^2$ is everywhere
positive if $v < v_-(\rho_0)$ or $v > v_+(\rho_0)$, where
 \be
v_{\pm}(\rho_0) = \dfrac{1\pm\sqrt{1-\beta^2}}2\,\rho_0^{2q-1}\,,
 \ee
leading to two disjoint sectors of regular black holes. In the third
parameter sector $v_-(\rho_0) < v < v_+(\rho_0)$, there are CTC which,
according to the above argument, are naked.

However, the absence of naked curvature and causal singularities are
not enough to guarantee a regular black hole. One must also check
that geodesics do not terminate at spatial infinity. The first
integral for the geodesics in the metric (\ref{an}) is \cite{C94}
 \be \lb{geo}
\left( \frac{d\rho}{d\tau} \right)\!^2 + \mathbf{P}.\mathbf{X} +
\varepsilon R^2 = 0\,,
 \ee
where $\tau$ is an affine parameter, $\mathbf{P}$ a constant future
null vector, and $\varepsilon = +1, 0$ or $-1$ for timelike, null,
or spacelike geodesics, respectively. For $q < 0$, Eq. (\ref{geo})
can be approximated for large $\rho$ by
 \be\lb{geoinf}
\left( \frac{d\rho}{d\tau} \right)\!^2  \simeq -
\mathbf{P}.\c\,\rho^{1-q}\,.
 \ee
Using the fact that the null vector $\c$ is, from ${\rm
sign}[k^{-1}(z+u)] = \epsilon'{\rm sign}(1-\beta^2)$, future in the
present case, the right-hand side of (\ref{geoinf}) is generically
positive. It follows that for $-1 \le q < 0$, geodesics extend to
infinity and the black hole metric is regular. The Penrose diagram
is similar to that of the Schwarzschild black hole (Fig. 1).

On the other hand, for $q < -1$ almost all geodesics terminate at
$\rho\to+\infty$. This is not a curvature singularity (the Ricci
scalar (\ref{riscal}) is finite), but a second horizon through which
the metric cannot be generically be extended. However, as in the
case of other ``cold black hole'' solutions of gravitating field
theories with negative gravitational constant in three \cite{sig} or
four \cite{cold,phantom} dimensions, characterized by an infinite
horizon area and vanishing Hawking temperature, the metric and the
geodesics can be analytically continued across the horizon for
discrete values of the model parameters. In the present case, Eq.
(\ref{geoinf}) suggests transforming to the radial coordinate $x =
\rho^{(1+q)/2}$, in terms of which the geodesic equation (\ref{geo})
can be written
 \be\lb{geo1}
\frac4{(q+1)^2}\left( \frac{dx}{d\tau} \right)\!^2 + \mathbf{P}
.\left(\c + \b\,x^m + \a\,x^{2m}\right) - \varepsilon\beta^2\,
x^n\left(\rho_0^{2q-1}-x^m\right)  = 0\,,
 \ee
where we have put
 \be
n = \frac{2q}{q+1}\,, \quad m = \frac{2(2q-1)}{q+1} = 3n-2\,.
 \ee
It is clear that Eq. (\ref{geo1}) is analytic, and therefore
geodesics can be extended from $x>0$ to $x<0$, if $n$ is an integer,
$n>2$, corresponding to the quantization condition
 \be
q = -\frac{n}{n-2}\,.
 \ee
Near the cold horizon $x=0$, the lapse function in (\ref{adm}) is
proportional to $x^n$, showing that this is a multiple horizon
(hence its vanishing Hawking temperature) of multiplicity $n$.

For \underline{$n$ odd}, $\rho=x^{2-n}$ and $\rho^{2q-1}= x^{3n-2}$
change sign when the cold horizon is crossed from the Lorentzian
region $I$ ($0<x<\rho_0^{-1/(n-2)}$) to the inner region $III$
($x<0$) bounded by the spacelike singularity $x=-\infty$ ($\rho=0$).
The resulting Penrose diagram is an infinite spacelike strip (Fig.
2). Using the fact that $\a$ and $\c$ are both future null, one can show that
almost all timelike or null geodesics originate from the past
spacelike singularity, cross successively the two horizons, and end
at the future spacelike singularity. Exceptional timelike or null
geodesics (those fine-tuned so that $\mathbf{P} = c\c$ with $c>0$)
do not cross the cold horizon. Exceptional geodesics with $x>0$ are
time-symmetric, originating from the past singularity and ending at
the future singularity after crossing twice the horizon
$\rho=\rho_0$. Exceptional geodesics with $x<0$ either originate
from the past singularity and asymptote the cold horizon, or follow
the time-reversed history.

For \underline{$n$ even}, analytic extension from $x>0$ to $x<0$
leads from the Lorentzian region $I$ to an isometric Lorentzian
region $I'$, so that the corresponding Penrose diagram paves the
whole plane (Fig. 3). The effective potential in (\ref{geo1}) is now
symmetrical in $x$. Typical timelike or null geodesics either join a
past singularity to a future singularity, without crossing, or
crossing only once the cold horizon $x=0$, or cross periodically the
cold horizon and extend to infinity.

To summarize, the solution (\ref{genmet}) or (\ref{adm}) (with
$k=\epsilon=-1$ and $\epsilon'=+1$) and (\ref{genA}) corresponds to
two disjoint sectors of regular black holes if the model parameters
are such that $\k < 0$, $\Lambda > \mu^2/4$ (ensuring from
(\ref{valb2}) that $\beta^2<0$ for $q < 0$), $b=-2c$, and either
$c^2 \ge -16\k$ (ensuring from (\ref{kq}) that $-1\le q<0$), or $c^2
= -16\k(n-1)/n$ ($q = -n/(n-2)$) with $n>2$.

\setcounter{equation}{0}
\section{Mass, angular momentum and thermodynamics}
The metric (\ref{adm}) is neither asymptotically flat nor
asymptotically AdS, so that neither the ADM approach \cite{ADM} nor
the Abbott-Deser-Tekin (ADT) approach \cite{AD,DT02}, which involve
linearization around a flat or constant curvature background, can be
used to compute the conserved black hole charges, which are its mass
and angular momentum. In \cite{BC} the ADT approach was extended to
compute the conserved charges of a solution of topologically massive
gravity linearized around an arbitrary background. This approach was
further generalized in \cite{ACGL} to the case of TMGE, and in
\cite{Nam} to that of new massive gravity. However, the authors of
\cite{Nam} pointed out that for warped AdS$_3$ black holes the
angular momentum obtained in these approaches was quadratic rather
than linear in the black hole parameters, so that, while the mass
and angular momentum obtained satisfied the first law of black hole
thermodynamics and so appeared to be correct, there was a problem
with the self-consistency of their derivation.

The solution to this problem was given recently in \cite{KKY},
generalizing an approach proposed in \cite{barnich}. Warped AdS$_3$
black holes --- and this is also true for the dilatonic black holes
considered here --- depend on two arbitrary parameters (integration
constants). Their asymptotics are such that their metric cannot be
considered to match at spacelike infinity that of any arbitrarily
given background solution, except one which is infinitesimally near
in parameter space to the solution under consideration. Following
the generalized ADT approach of \cite{BC}, one can compute the
corresponding differential conserved charges, from which the finite
conserved charges can be obtained as line integrals in parameter
space.

The differential mass and angular momentum of our dilatonic black
hole solutions are the Killing charges, defined as integrals  over
the boundary $\partial M$ of a spacelike hypersurface $M$
 \be
\delta Q(\x) = \frac1{\k}\int_{\partial M}\sqrt{|g|}\delta{\cal
F}^{0i}(\x)dS_i\,,
 \ee
of the superpotentials
 \be\lb{supots}
\delta{\cal F}^{\mu\nu}(\x) = \delta{\cal F}^{\mu\nu}_g(\x) +
\delta{\cal F}^{\mu\nu}_e(\x)
 \ee
associated with the Killing vectors $\x = \partial_t$ and $\x =
\partial_{\varphi}$. In (\ref{supots}) the gravitational contribution
is the Einstein superpotential given in \cite{BC} (Eq. (2.14)), and
the electromagnetic contribution is obtained from that of
\cite{BBCG} by replacing the fields canonically conjugate to the
vector potential $A$ by $e^{c\phi}F$,
 \ba\lb{Fe}
\delta{\cal F}^{\mu\nu}_e(\x) &=&
\frac{\k}{\sqrt{|g|}}\delta\bigg[\sqrt{|g|} e^{c\phi}F^{\mu\nu} -
\mu\epsilon^{\mu\nu\lambda}A_{\lambda}\bigg]\xi^{\rho}A_{\rho} \nn
\\ && + \k e^{c\phi}\bigg[F^{\mu\nu}\xi^{\rho} +
F^{\nu\rho}\xi^{\mu} +  F^{\rho\mu}\xi^{\nu}\bigg]\delta A_{\rho}\,.
 \ea
For a rotationally symmetric configuration, the relevant component
is $\delta{\cal F}^{02}_e(\xi)$. The first bracket of (\ref{Fe})
vanishes by virtue of (\ref{firstpsi}), and there remains
 \be
\delta{\cal F}^{02}_e(\x) = -\k F^{2a}\epsilon_{ab}\xi^b\delta
\psi_1 = -\k\zeta^2(\xi^T\psi)\delta\ol\psi^0 =
\frac{\zeta^2}2\bigg[\xi^T\mathbf{\Sigma}.\delta\S -
\k(\delta\ol{\psi}\psi)\xi^T\bigg]^0\,,
 \ee
as in \cite{ACGL}. Therefore, the Killing charges are obtained from
those given there by omitting the gravitational Chern-Simons term:
 \begin{equation}
\delta Q(\x )=\frac{\pi \zeta }{\kappa }\left[ \xi ^{T} \left(
\mathbf{\Sigma}. \delta \mathbf{J}+\Delta \right) \right] ^{0}\,,
\end{equation}
where $\mathbf{J}$ is the super angular momentum \cite{tmge},
 \be
\J = \X\wedge\X' + \S\,,
 \ee
which is constant by virtue of (\ref{eqX}) and (\ref{eqS}), and
$\Delta$ is the scalar
\begin{equation}
\Delta = \mathbf{X}.\delta \mathbf{X}^{\prime }-\kappa \left( \delta
\overline{\psi }\psi \right)\,.
\end{equation}
For our black hole solutions we obtain, from (\ref{an}),
(\ref{Sabc}) and (\ref{kq}) with $k=-1$,
 \be\lb{Ja}
\J = (1-2q)[\b\wedge\c  + 2(\a.\c)\c]\,,
 \ee
and,
 \be\lb{Da}
\Delta = (1-2q)(2-\beta^2)\,(\b.\delta\c) + q\delta(\b.\c) =
-\beta^2\,\delta\left[(1-2q)v + \frac{q}2\rq\right]\,.
 \ee
We note here that, had we defined $\delta\X'$ not as a differential,
but as the finite (linearized) difference $\X'-\X'_0$, with $\X_0$ a
given background solution, $\Delta$ would have contained in addition
a non-constant contribution proportional to
$(\c.(\c-\c_0))\rho^{1-2q}$, which is absent here because
$(\c.\delta\c) = 0$ by virtue of the constraint (\ref{cnull}). So in
the present case the definition of the Killing charges as
differentials is essential to guarantee their conservation.

The black-hole mass and angular momentum are respectively the
Killing charges for the vectors $\x =(-1,0)$ and $\x =(0,1)$,
 \ba
M &=&-\frac{\pi\mu}{\kappa}\left( \delta J^{Y}+\int\Delta
\right)\,,\nn\\
J &=& \frac{\pi\mu}{\kappa}\left( \delta J^{T}-\delta
J^{X}\right)\,,
 \ea
where the integral over $\Delta$ is a line integral from the
``vacuum'' solution $v=\rq=0$ to the solution under consideration.
We obtain from (\ref{Ja}) and (\ref{Da}):
 \ba
M &=& \frac{\pi\mu\beta^2}{\kappa }\left[2(1-2q)v -
\frac{1-3q}2\rq\right]\,,\lb{M}\\
J &=& \frac{\pi\mu\beta^2}{\kappa
}\frac{1-2q}{1-\beta^2}\,v\left(v-\rq\right)\,.
 \ea
Noting that for our black holes $\k<0$ and $\beta^2<0$, we find that
the mass $M$ is positive in the black-hole sector $v>v_+$ ($>\rq$),
and negative in the black-hole sector $v<v_-$ ($<0$), while the
angular momentum $J$ is positive in both sectors.

A third black-hole observable is its entropy, proportional to its
horizon areal radius $r_h$ (computed in Appendix A),
 \be
S=\frac{4\pi^{2}}{\kappa}r_{h} =
\frac{4\pi^2}{\k\sqrt{1-\beta^2}}\,\rho_0^{(1-q)/2}|v-\rq|\,,
 \ee
which is negative. The other thermodynamical observable is the
Hawking temperature,
\begin{equation}
T = \frac{\mu(R^2)'(\rho_0)}{4\pi r_h} =
-\frac{\mu\beta^2\sqrt{1-\beta^2}(1-2q)}{4\pi}\,\frac{\rho
_0^{(5q-3)/2}}{|v-\rq|}\,.
\end{equation}
Finally, the horizon angular velocity is
\begin{equation}
\Omega _{h} = -\frac{Y(\rho_0)}{r_h^2} = \frac{1-\beta^2}{v-\rq}\,.
\end{equation}
This is positive in the sector $v>v_+$ (co-rotating black holes) and
negative in the sector $v<v_-$ (counter-rotating black holes). Let
us recall that counter-rotating black holes (whose horizon rotates
in the opposite sense to the angular momentum) have previously been
numerically constructed in four-dimensional Einstein-Maxwell-dilaton
theory (with dilaton coupling constant larger than $\sqrt3$)
\cite{kunz1}, and in five-dimensional Einstein-Maxwell-Chern-Simons
theory (with Chern-Simons coefficient larger than $1$) \cite{kunz2}.

These values can be checked to be consistent with the first law of
black hole thermodynamics for independent variations of the black
hole parameters $v$ and $\rho_0$,
\begin{equation}\lb{first}
dM=TdS+\Omega_{h}dJ\,.
\end{equation}
We also note that the Chern-Simons dilaton black holes satisfy the
integral Smarr-like relation
\begin{equation}
M=\frac{1-3q}{2(1-2q)}TS+2\Omega _{h}J \,,
\end{equation}
which generalizes that given in \cite{ACGL} for Chern-Simons black
holes ($q=1$).

\setcounter{equation}{0}
\section{Solution of five-dimensional Einstein-Maxwell-Chern-Simons theory}
Dimensional reduction of higher-dimensional field theories
generically leads to the appearance of dilaton fields, so one can
surmise that, at least in a domain of parameter space, the theory
(\ref{act}) results from the reduction of some higher-dimensional
field theory. In \cite{reds2}, five-dimensional
Einstein-Maxwell-Chern-Simons theory was reduced, by monopole
compactification on a constant curvature two-surface, to
three-dimensional Einstein-Maxwell-Chern-Simons theory with an
additional constraint. This ``hard'' reduction led to
five-dimensional solutions which included a class generated from the
three-dimensional warped AdS black holes. We now show that a special
``soft'' reduction, where the constraint is avoided by introducing a
dilaton field, leads to a subclass of the theory (\ref{act}).
Conversely, oxidation of a subclass of the solutions derived in the
present paper therefore leads to non-asymptotically flat solutions
of the five-dimensional theory.

Five-dimensional Einstein-Maxwell-Chern-Simons theory is defined by
the action
 \be\lb{emcs5}
S_5 = \frac1{16\pi G_5}\int d^5x \bigg[\sqrt{|g_{(5)}|} \bigg(R\5 -
\frac14F\5^{\mu\nu}F_{(5)\mu\nu}\bigg)    -
\frac\gamma{12\sqrt3}\epsilon^{\mu\nu\rho\sigma\lambda}F_{(5)\mu\nu}
F_{(5)\rho\sigma}A_{(5)\lambda} \bigg]\,,
 \ee
where $F\5 = dA\5$, $\mu,\nu,\cdots = 1,\cdots,5$, and $\gamma$ is
the Chern-Simons coupling constant, the value $\gamma=1$
corresponding to minimal five-dimensional supergravity. Let us
assume for the five-dimensional metric and the vector potential the
warped product ans\"atze
 \ba\lb{softred}
ds\5^2 &=&
e^{-c\phi(x^{\gamma})}g_{\alpha\beta}(x^{\gamma})dx^{\alpha}dx^{\beta}
+ e^{c\phi(x^{\gamma})/2}\,a^2(dx^2+dy^2)\,, \nn\\ A\5 &=&
\sqrt{2\k}\,A_{\alpha}(x^{\gamma})dx^{\alpha} + \frac{e}2\,(xdy -
ydx)\,,
 \ea
where $\alpha,\beta,\gamma=1,2,3$, $x^4=x$, $x^5=y$ are the
transverse space coordinates, and $a>0$ is an arbitrary scale. The
constant transverse magnetic field $F_{xy} = e$ solves identically
the corresponding Maxwell equations, and its contribution to the
five-dimensional energy-momentum tensor is proportional to the flat
transverse space metric. The remaining five-dimensional field
equations reduce to three-dimensional equations deriving from the
action (\ref{act}), with $b=-2c$ and the identifications
 \be
\Lambda = \frac{e^2}{4a^4}\,, \quad c = 4\sqrt{\frac{2\k}3}\,, \quad
\mu = \frac{2\gamma e}{\sqrt3a^2}\,.
 \ee

These equations are solved by (\ref{ecphi}), (\ref{genmet}) and
(\ref{genA}) with
 \be
q = \frac37\,, \quad \beta^2 =
-\frac12\left(1-\frac3{4\gamma^2}\right)\,, \quad k =
-7\rho_1^{-4/7}\,,
 \ee
Because in the present case $\k$ is positive and $q<1/2$, it follows
from the analysis of Sect. 4 that this is a stationary solution
($g_{\rho\rho}$ is positive for $\rho>\rho_0$) if $\beta^2 < 0$,
i.e. $\gamma > \sqrt3/2$. Carrying out the coordinate transformation
 \ba
t &=& \sqrt{\dfrac7{1-\beta^2}}\,\tau\,, \quad \rho = (\alpha
r)^7\,, \nn\\
\varphi &=&
\dfrac{3\alpha}{7(4\gamma^2-3)}\sqrt{\dfrac{8(4\gamma^2-1)}7}
\dfrac{\rho_1^{8/7}r_0^3}e\,\psi\,,
 \ea
with $\alpha>0$ arbitrary, and fixing $r_1=a$, we find that this
three-dimensional solution oxidizes according to (\ref{softred}) to
the solution of the five-dimensional theory (\ref{emcs5})
 \ba\lb{sol5}
ds\5^2 &=& -\left[d\tau + \left(w -
\frac{br_0}r\right)d\psi\right]^2 + r^2(dx^2+dy^2) \nn\\ && +
\frac{\ol\beta^2b^2}{r_0} \left[\frac{r^2dr^2}{4r_0^2(r-r_0)} +
\frac{r_0^2(r-r_0)}{r^2}\,d\psi^2\right] \quad \left(b =
\frac{\sqrt3r_0^2}{\gamma e\ol\beta^2}\right)\,,\nn\\
A\5 &=& \pm\sqrt{2(1+\ol\beta^2)}\left[d\tau + \left(w -
\frac{br_0}r\right)d\psi\right] + \frac{e}2\,(xdy-ydx)
 \ea
(with $\ol\beta^2 = -\beta^2$) depending on three parameters $e$,
$r_0$ and $w$. This solution has a bolt singularity at $r=r_0$, but
becomes regular if $\psi$ is an angle (with period $2\pi$). Then the
range of $r$ is $r>r_0>0$, so that the singularity at $r=0$ becomes
irrelevant.

One could expect the ``soft'' solution (\ref{sol5}) to reduce to the
``hard'' G\"{o}del class solution of \cite{reds2} in the near-bolt
limit $r\to r_0$, where the dilaton field can be replaced by a
constant. This turns out not to be possible because, for $\beta^2<0$
($\gamma > \sqrt3/2$), the constant curvature transverse two-spaces
of the G\"{o}del class solutions are not flat but hyperbolic (with
negative curvature). However we show in Appendix B that there is
indeed a connection, the ``double near-bolt'' limit of the hard
solution coinciding with the near-bolt limit of the soft solution.

\section{Conclusion}
We have constructed rotating magnetic solutions to the
three-dimensional Einstein-Maxwell-Chern-Simons-dilaton theory
defined by the action (\ref{act}) with $b=-2c$. These include for
$\k < 0$ and $\Lambda > \mu^2/4$ a class of black hole solutions,
which generalize the warped AdS black holes of \cite{ACGL}. In the
range $c^2 > -16\k$ of the dilaton coupling constant $c$, the
regular black holes are Schwarzschild-like, with a horizon shielding
a spacelike singularity. For the discrete values $c^2 =
-16\k(n-1)/n$ (with $n>2$), null infinity is replaced by a second
multiple horizon with vanishing Hawking temperature. We have also
computed the mass, angular momentum and other thermodynamic
observables for these solutions. The regular black holes belong to
two disjoint sectors. The black holes of the first sector have a
positive mass and are co-rotating, while the black holes of the
second sector have a negative mass and are counter-rotating.

We have also shown that a particular, non-black hole, subfamily of
our three-dimensional solutions may be uplifted to new regular
non-asymptotically flat solutions of five-dimensional
Einstein-Maxwell-Chern-Simons theory. We conjecture that it might be
possible to generalize such a construction, both by taking into
account a five-dimensional cosmological constant, and by extending
the reduction ansatz (\ref{softred}) to a constant curvature
transverse two-space, provided that the solution-generation
technique of the present paper could be extended to the case of a
dilaton potential more general than the Liouville potential in
(\ref{lag}).

Returning to the three-dimensional theory (\ref{act}), we comment on
our apparently arbitrary choice of the relation $b + 2c = 0$ between
the dilaton couplings. We would not expect non-trivial closed-form
solutions for dilaton couplings not satisfying this constraint,
which might lead to a symmetry enhancenent of the theory. In this
respect, the situation might be similar to that of four-dimensional
Einstein-Maxwell-dilaton theory, which admits closed form rotating
solutions only for specific values of the dilaton coupling constant
$\alpha$ ($\alpha = 0$ or $\alpha = \sqrt{3}$), which are precisely
the values for which the symmetries of the theory are enhanced
\cite{emd4}. Whether this is indeed the case for our
three-dimensional theory remains to be investigated.

\subsection*{Acknowledgments}
We acknowledge discussions with Adel Bouchareb at an early stage of
this work. KAM and HG would like to thank LAPTh Annecy-le-Vieux for
hospitality at different stages of this work. KAM acknowledges the
support of the Ministry of Higher Education and Scientific Research
of Algeria (MESRS) under grant D00920140043.

\renewcommand{\theequation}{A.\arabic{equation}}
\setcounter{equation}{0}
\section*{Appendix A}
From (\ref{mink}),
 \be\lb{Ar2}
r^2 = g_{\varphi\varphi} = \left[-g_{tt}\right]^{-1}(Y^2-R^2)\,,
 \ee
where, for the metric (\ref{adm}) with $\kappa < 0$ ($q<0$,
$\epsilon'=1$, $\epsilon=-1$) and $\beta^2<0$,
 \be
g_{tt} = (1-\beta^2)\rho^{1-q} > 0\;, Y =
\rho^{1-q}\left(\rho^{2q-1}-v\right)\;, R^2 =
-\beta^2\rho\left(\rho_0^{2q-1}-\rho^{2q-1}\right)\,.
 \ee
Putting $x = \rho^{2q-1}$, this leads to
 \be
r^2 = \frac{\rho^{1-q}}{1-\beta^2}\left[(1-\beta^2)x^2 -
(2v-\beta^2x_0)x + v^2\right]\,.
 \ee

The discriminant of the quadratic polynomial in brackets
 \be
\Delta = -4\beta^2\left[-v^2 + vx_0 - \beta^2x_0^2\right]
 \ee
has two roots
 \be
v_{\pm}(x_0) = \left[1\pm\sqrt{1-\beta^2}\right]\frac{x_0}2\,.
 \ee
For $v_- < v < v_+$, the discriminant is positive, so that $r^2$ has
two roots $r_\pm$ which are both of the same sign as
 \be
2v-\beta^2x_0 > 2v_--\beta^2x_0 = \left[\sqrt{1-\beta^2}-1\right]x_0
> 0.
 \ee
These two roots correspond to two causal singularities of the metric
(\ref{adm}), which are naked because $R^2(x_\pm) = Y^2(x_\pm) > 0$.
Thus, this metric leads to regular black holes only if either $v >
v_+$, or $v<v_-$. From (\ref{Ar2}) the corresponding horizon radii
are
 \be
\begin{array}{lccc}
v>v_+\,: & r_h = \dfrac{\rho_0^{(1-q)/2}}{\sqrt{1-\beta^2}}\,(v-x_0)
&
> &
\dfrac{\rho_0^{(3q-1)/2}}{\sqrt{1-\beta^2}}\,(\sqrt{1-\beta^2}-1)\,, \\
v<v_-\,: & r_h = \dfrac{\rho_0^{(1-q)/2}}{\sqrt{1-\beta^2}}\,(x_0-v)
& > &
\dfrac{\rho_0^{(3q-1)/2}}{\sqrt{1-\beta^2}}\,(\sqrt{1-\beta^2}+1)\,,
\end{array}
 \ee
where we have used $v_+ > x_0 > 0$ and $v_- < 0 < x_0$.

\renewcommand{\theequation}{B.\arabic{equation}}
\setcounter{equation}{0}
\section*{Appendix B}
The ``G\"{o}del class'' solution (3.22) of \cite{reds2} for
$\beta^2<0$, with $k=-1$ (negative constant curvature transverse
two-space) and vanishing five-dimensional cosmological constant
($\ol{g} = g = a\ol\beta^{-1}$) is
 \ba\lb{bgod3}
ds\5^2 &=& - \left(dt - g\cos\chi\,d\psi\right)^2
+g^2\ol\beta^2\left[d\chi^2 + \sin^2\chi\,d\psi^2 + d\theta^2 +
\sinh^2\theta\,d\varphi^2\right] \,,\nn\\
A\5 &=& \pm\sqrt{2(1+\ol\beta^2)}\left(dt - g\cos\chi\,d\psi\right)
+ \ol{e}\,\cosh\theta\,d\varphi \quad (g = 2\gamma\ol{e}/\sqrt3)\,.
 \ea
In the double near-bolt limit $\chi \to 0$, $\theta \to 0$, this
goes to
 \ba\lb{limbgod3}
ds\5^2 &\simeq& - \left(d\ol{t} + \frac{g}2\chi^2\,d\psi\right)^2
+g^2\ol\beta^2\left[d\chi^2 + \chi^2\,d\psi^2 + d\theta^2 +
\theta^2\,d\varphi^2\right] \,,\nn\\
A\5 &\simeq& \pm\sqrt{2(1+\ol\beta^2)}\left(d\ol{t} +
\frac{g}2\chi^2 \,d\psi\right) + \frac{\ol{e}}2\,\theta^2\,d\varphi
\,,
 \ea
with $\ol{t} = t-g\psi$.

On the other hand, the near-bolt limit $r \to r_0$ of the solution
(\ref{sol5}) yields
 \ba\lb{limsol5}
ds\5^2 &\simeq& - \left(d\ol{t} + \frac{b}4\chi^2\,d\psi\right)^2
+\frac{b^2\ol\beta^2}4\left[d\chi^2 + \chi^2\,d\psi^2 + d\theta^2 +
\theta^2\,d\varphi^2\right] \,,\nn\\
A\5 &\simeq& \pm \sqrt{2(1+\ol\beta^2)}\left(d\ol{t} +
\frac{b}4\chi^2 \,d\psi\right) +
\frac{eb^2\ol\beta^2}{8r_0^2}\,\theta^2\,d\varphi \,,
 \ea
where we have put $\ol{t} = \tau + (w-b)\psi$, and
$$r \simeq r_0(1+\chi^2/4)\;\; (\chi\to0)\,, \quad x =
\frac{b\ol\beta}{2r_0}\theta\cos\varphi\,, \quad y =
\frac{b\ol\beta}{2r_0}\theta\sin\varphi\,.$$ The two limits
(\ref{limbgod3}) and (\ref{limsol5}) coincide if $g = b/2$, leading
from the definitions of $g$ in (\ref{bgod3}) and of $b$ in
(\ref{sol5}) to $\ol{e} = (b^2\ol\beta^2/4r_0^2)e$.

\newpage

\begin{figure}
\centering
\includegraphics[scale=0.7]{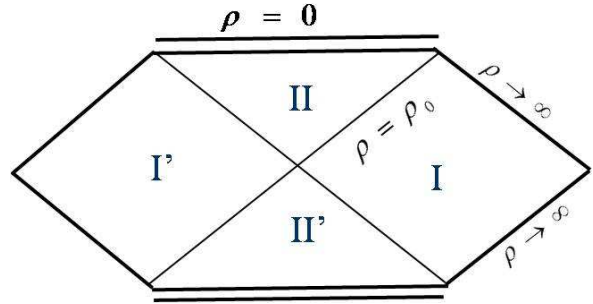}
\caption{Penrose diagram of the black holes with $-1\le q<0$.}
\end{figure}

\begin{figure}
\centering
\includegraphics[scale=0.7]{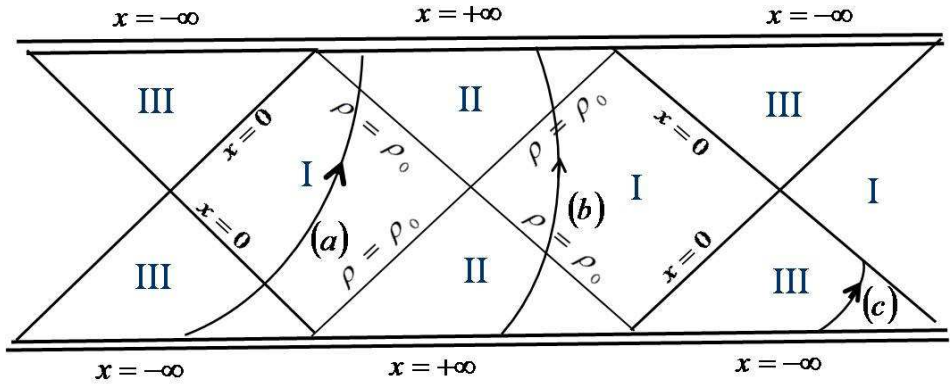}
\caption{Penrose diagram of the cold black holes with $n$ odd; ($a$)
is a normal geodesic, ($b$) and ($c$) are exceptional geodesics.}
\end{figure}

\begin{figure}
\centering
\includegraphics[scale=0.7]{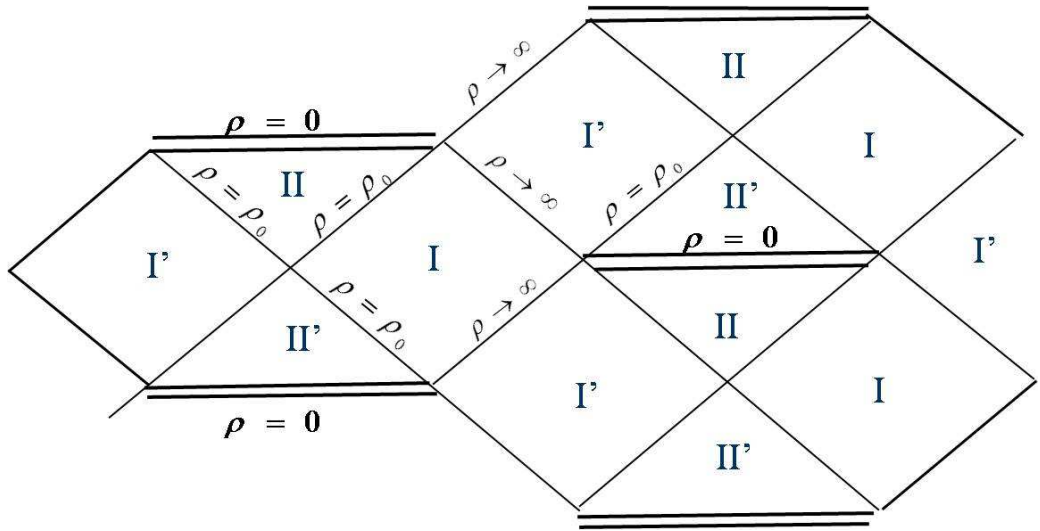}
\caption{Penrose diagram of the cold black holes with $n$ even.}
\end{figure}

\end{document}